\documentstyle[12pt,epsf]{article}
\begin{document}

\title{\bf Physical Chirality.\\
  It Feeds on Negative Entropy
  \thanks{Published at {\bf Fundamentals of Life} \newline \copyright \ 2002 \'{E}ditions scientifiques et
  m\'{e}dicales Elsevier SAS. All rights reserved \newline
  Eds., G. Palyi, C. Zucchi, L. Caglioti. \newline
  3-8 Sep. 2000, Modena, Italy.}}
\author{G. Gilat \\ \\
  Department of Physics \\
  Technion, Haifa 32000, Israel \\
  email: gilat@physics.technion.ac.il \\
}
\date{}
\maketitle
%
%\thanks{Presented and published at Fundamentals of Life
%  \copyright \acute{E}ditions scientifiques et m\acute{e}dicales Elsevier SAS. All rights reserved
%  Ed. G. Palyi, C. Zucchi, L. Caglioti, 3-8 Sep. 2000, Modena, Italy.}
%

\bigskip
\begin{abstract}
\noindent
Chirality is considered by many scientists to be mainly a
geometric concept. There exists also a physical aspect of
chirality which is largely being overlooked at.  Two examples of
mechanical devices are introduced here that represent ``Physical
Chirality''. \ These are a rotating water sprinkler and a variant
of Crookes' radiometer. When interacting with appropriate media,
they both choose only one mode of rotation out of two possible
ones. Such a behavior does not obey time-reversal invariance,
which is regarded to be a rule in classical mechanics. This is due
to their chiral nature. Instead, they do obey a space-time (ST)
law of invariance, that is, what is rotating in the opposite
direction is the mirror-image of the given device. In a recent
experiment of Koumura et al. they discovered a similar behavior of
a molecular rotor. The possible biological significance of
physical chirality is emphasized hereby, and the conclusion is
that chiral molecular systems do not reach readily thermal
equilibrium. In other words: ``Physical chirality does feed on
negative entropy'', and therefore, it may well be of crucial value
to life.
\end{abstract}
%
%\pagebreak
%\noindent {\bf I. Introduction} \\
%
\begin{section}{Introduction}
The phenomenon of structural chirality of crystals and molecules
has been recognized since the early 19th century when Arago$^1$ and Biot$^2$
did demonstrate the effect of optical activity in quartz crystals.
Louis Pasteur$^3$ was the first one to observe chirality on a
molecular level and specified it as ``dissymmetry''. The term
``Chirality'' was first introduced by Kelvin$^4$, who also defined this
concept as a property of any object that cannot superimpose, or
overlap, completely its own mirror image.

The main reason for Kelvin to modify this term came from the fact that
chirality did not necesssarily imply a total lack of symmetry.
Actually, chirality does lack the symmetry property of reflection,
but chiral shapes of bodies may also obey so-called continuous symmetry
operations such as rotational and/or translational symmetries.

At this point it becomes relevant to point out that the concept of
chirality in science is largely considered to be mainly of geometrical
nature. Even Kelvin's definition of chirality concerns mainly the
geometric shape of a body, regardless of its physical properties.
For this very reason it becomes necessary to point out possible
physical aspects of chirality and this becomes one of the main
motivations of the present article.  A physical body may contain in
it various different materials. It may also contain different properties
such as colors or magnetic structure.  All these are not considered
as realistic from geometric view point. Such properties may have
physical aspects and they form the basis for what is referred to as
``Physical chirality''. \ A body that contains this property of
physical chirality may operate in a perculiar mode that differs
considerably from a regular mode of operation of an achiral body.
The source of such an operation
comes from a specific
type of {\bf interaction} that exists between various mechanical devices
and different media such as flow of air or water and even light
radiation. What is special about this type of interaction is the presence
of chiral structure in these devices which makes their mode of
operation quite different from other interactions which are based
on achiral objects such as the Newtonian mass point. Such an
interaction is to be labeled ``chiral interaction'' (CI) and it has
already been described and treated in several publications.$^{5-7}$ \
It is important to note that this phenomenon of physical chirality is
largely being ignored and overlooked in classical mechanics, and many
physicists are quite unaware of its physical significance
in the domain of biophysics.
\end{section}
%
%\vspace{0.5cm}
%
%\noindent {\bf II \ Chiral Interaction in Mechanical Devices} \\
%
\begin{section}{Chiral Interaction in Mechanical Devices}
As mentioned above,
chiral interaction is not limited to molecular structure only, but there
exist various mechanical chiral devices that function according to the
same principle.
The most obvious chiral device is the rotating water sprinkler (see
Fig. 1). When the water stream enters the sprinkler it ``knows''
immediately in which direction to rotate. This is so due to its
chiral structure. What is rotating in the opposite direction is the
mirror image of the given sprinkler. Let us define
now by D and L the ``enantiomers'' of the rotating sprinkler according
to its direction of rotation being clock or anti clockwise, respectively.
Another common chiral device is the windmill. When wind blows at the
vanes of the mill it also ``knows'' instantly to which direction
to rotate. This is due to its chiral design. If it
did contain reflection
symmetry with respect to its axis of rotation, it ``would not know''
to which direction to rotate.
%
%\newpage
%
The next example, shown in Fig. 2, is somewhat
more sophisticated, and it depends on a different mode of chirality.
This device is a simple variant of the Crookes' radiometer. The active
medium in this case is light radiation. The element of chirality here
consists of two different colors on both sides of the
rotating blades, being black and silver, respectively. This is a special
example of a physical rather than of a geometric chirality.
From a pure geometric sense of view this radiometer contains a complete
reflection symmetry and, therefore, it can be regarded as achiral.
Physical chirality$^{7,8}$ is presented by the chiral distribution
of a physical property rather than by a chiral geometric shape.
Physical chirality differs from a geometric one in its capability
of interacting with various media surrounding it.
In the case of this special example of a variant of the ordinary Crookes'
radiometer, the physical distribution of the black and silver colors
on the blades represents a large difference in the light absorption
coefficient
of the blades. The silver side reflects back the light, whereas
the black side absorbs the light and therefore becomes
warmer in comparison to the silver
one. This causes the air close to the black side to become heated,
and as a result it expands and pushes back the black side,
which ends up in rotating the device
in the preferred direction
of the black side of the blade. The selection of
the sense of rotation of the blades is made
by the variance of colors on the blades and their interaction with
light. The concept of physical chirality$^{7-8}$ here is represented
by the distribution
of the optical absorption coefficient on the blades and {\bf not}
by their geometric shape. \\
%
%figure 1
%
%\pagebreak
%\noindent {\bf Figure Caption} \\
%
\begin{figure}[ht]
\centerline{\epsfysize=12cm \epsffile[80 170 540 715]{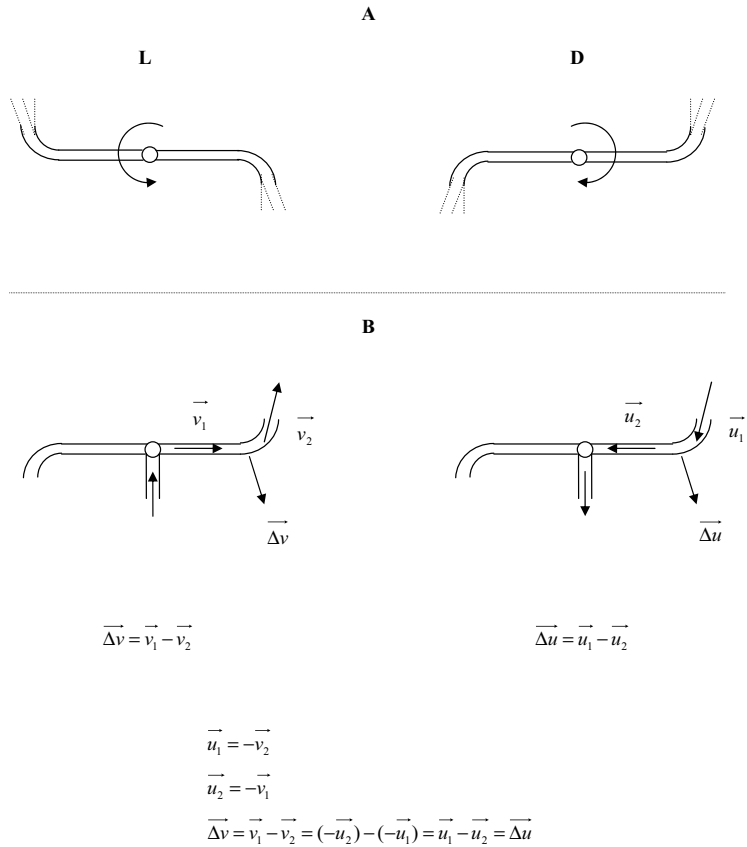}}
\caption{(A) \ Two rotating water sprinklers are shown as
examples of chiral devices. The sprinkler rotating in clock-wise
direction when looking at it from above is marked by D and its mirror
image by L. At the bottom, both sprinklers are shown from above. These
provide for an excellent example of time irreversible macro chiral
device. \ (B) \ At the bottom it is shown that by reversing
the direction of
the flow of water, the vectorial difference of the water flow
momentum at the
bending of the sprinkler, does still rotate the sprinkler
in the same direction.
This is why the rotating water sprinkler is space-time (ST)-invariant.}
\end{figure}
\newpage
One of the main objectives for carefully analyzing the action of chiral
mechanical device is to point out a certain unusual result that is
derived from their operation. The choice of only one direction of
rotation by these devices is not readily acceptable in classical
mechanics.  Actually, the meaning of this mode of operation is that
chiral interaction does not obey time-reversible invariance. By
reversing the direction of motion of the interacting media, such as
the flow of water in the sprinkler (see Fig. 1), or the direction of
the light radiation in the case of the radiometer (see Fig. 2), the
direction of rotation of the devices
does not change, it
remains the same. This is the
main effect of chirality in these cases. What does rotate in the
opposite direction is the mirror-image of each of these chiral devices.
This effect happened also to irritate Richard Feynman who tried to
reverse the direction of flow of water in the sprinkler and caused
a flood in his laboratory at Los Alamos.  Actually, by looking
carefully at the bending of the sprinkler, where the direction of the
momentum of the water is varying and it creates an angular momentum to
the sprinkler with respect to its axis. It can now be readily deduced
that the difference in the water momentum will create the same direction
of angular momentum, regardless of which side the water is flowing
from (see Fig. 1). \\
%
% figure 2
%
\begin{figure}[htb]
\centerline{\epsfxsize=10cm \epsffile[130 220 480 560]{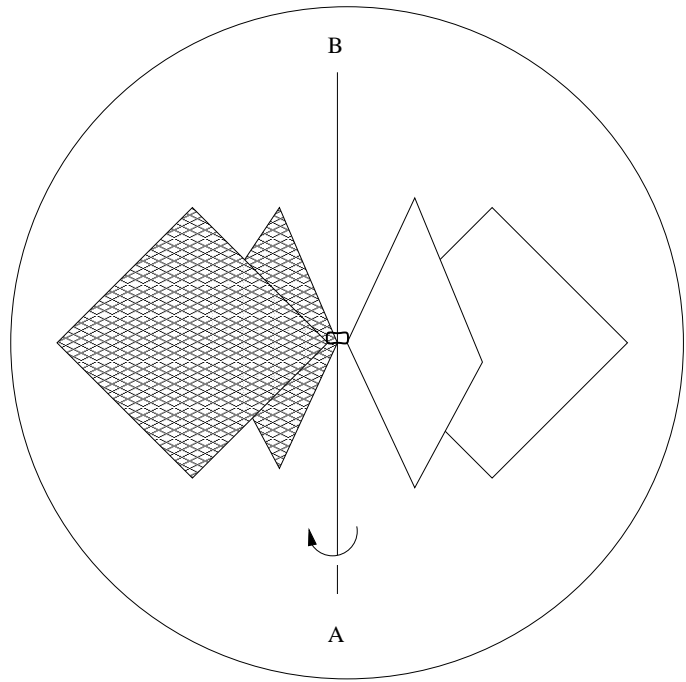}}
\caption{A variant of Crookes' radiometer is an example
of the operation of chiral interaction (CI). The asymmetry in the optical
absorption coefficient
between the black and the silver blades generates a temperature difference
between them when light is shining at the device. This expands the air
on the side of the black blade which, in turn, pushes it around
the axis AB in
the preferred direction towards the black vane. This is an
example of a physical
rather than of a geometric chirality where the chirality is in the
colors of the blades and not in their geometric shape.}
\end{figure}
\newpage
A similar conclusion is derived from the action
of the radiometer. It always rotates in the direction of the black
blade regardless of the direction from which the light is shining
at. In what follows this phenomenon is further treated on.

To summarize the main features of chiral interaction (CI) in
mechanical-devices, let us notice
that in all these examples there exists a specific medium with which
the chiral device is interacting and this is always happening
at an interface
separating the device from the active medium. The physical chirality is
{\bf built} into this very interface. CI is a process by which energy
is transferred from the active medium into the chiral device
which causes a rotational motion, being usually of mechanical
nature. The most significant aspect of the chiral
interaction process is
its mode of selecting only {\bf one} direction of rotation out of
{\bf two} possible ones, which is to be attributed to the {\bf chiral}
nature of the device. The {\bf mirror image} of the given chiral device,
interacting with the same medium, does produce the same rotational motion
in the {\bf opposite} direction. This is to be regarded as a main
feature of chiral interaction.

The effect of chiral interaction (CI) on a molecular level is less
recognizable in comparison to that of macro-chiral devices. The
main reason for this is that CI occurs mainly within the chiral
system, or molecule, in the form of rather small perturbations
which are not easy to detect experimentally. In a recent
experiment by Koumura et.al$^9$ they observed also a similar
behavior of chiral interaction on a molecular level of
``monodirectional molecular rotor''. \ Much more recognizable are
the physical effects associated with molecular chiral structure,
such as optical activity and related effects. These are to be
regarded as ``chiral scattering'', rather than CI, since the
observable effect concerns the polarized light being scattered
away from the chiral molecule, rather than its effect on the
molecule itself which is intrinsic and cannot be readily
detectable. This is chiral interaction$^{7}$.

A physical model of chiral interaction (CI) in soluble proteins
and amino-acids has already been developed and described in detail
in several\\ publications$^{5-7,10-12}$. The description here
contains only a few main features of this model.  The active
medium in this model consists of a random motion of ions,
including protons, throughout the solvent, being mostly regular
water. The chiral element that interacts with these ions is an
electric dipole moment that exists in the protein structure. This
interaction causes a moving ion to be deflected away from its
original track of motion, which creates a continuous perturbation
along the $\alpha$-helix of which the proteins consist, and this
perturbation moves along the helix in {\bf one} preferred
direction out of {\bf two} possible ones. Such a perturbation of
each ion is energetically rather small, but they add up together
for the whole protein molecule and become comparable to the size
of thermal energy. This is an abbreviated description of the model
of CI that occurs in soluble proteins. A more detailed description
appears in earlier publications.$^{5-7,10-12}$ The perturbation
resulting from this CI is of electric nature, rather than a
mechanical one. Another interesting aspect of this CI is that it
is happening at an interface separating the interior of the
protein molecule from the solvent. This is due to the globular
structure of the soluble protein. It is well known that all
soluble proteins become globular before they can function as
enzymes$^{13}$.

As mentioned above chiral interaction is not easy to observe
experimentally on a molecular level due to the small size of this
effect. Nevertheless, there exists a certain strong supporting
evidence for its existence owing to an experiment performed by
Careri et al.$^{14}$. This experiment concerns the effect of
dehydration on the protonic, or ionic, motion throughout the
hydration layers surrounding soluble proteins. The amount of water
around each protein is crucial for free protonic motion around the
molecule, which is also crucial for chiral interaction. By
dehydrating these water layers, a level is being reached when
protonic motion becomes awkward and stops, and so does also,
simultaneously, the enzymatic activity of the protein molecule. On
re-hydrating the molecule, protonic motion becomes possible again
and this, in turn, causes also the onset of enzymatic activity of
the protein molecule. This experiment shows$^{14}$ that free ionic
motion around soluble protein molecules, which is a necessary
condition for chiral interaction, is also necessary for their
enzymatic activity.
\end{section}
%
%\vspace{0.5cm}
%
%\noindent {\bf III Physical Aspects and Biological Significance} \\
%
\begin{section}{Physical Aspects and Biological Significance}
The main objective of the present article is to draw several physical
conclusions from the phenomenon of chiral interaction
in macro-chiral devices,
which are quite different from the
regular rules that exist in classical mechanics. The source of these
differences arises from the presence of physical chirality
as a major
feature, instead of the achiral environment that plays a basic role in
classical mechanics.
From these conclusions analogies
can be drawn for the function of molecular chiral systems, which may
well be of considerable significance in molecular biology.

The first conclusion concerns the symmetry operation of time-reversibility
that exists in many examples of classical physics.
As mentioned before, in the case of chiral
interaction time-reversibility cannot be conserved. The
rotating water sprinkler,
rotates about its axis in a single preferred direction due to
its chiral design.
What is rotating in
the opposite direction is the {\bf mirror-image}
of the given sprinkler.
The meaning of this mode of symmetry operation is that a
rotating water sprinkler
is not time-reversible, but it obeys space-time invariance.
The meaning of this conclusion is not very common in classical
mechanics. The reason for this is that the physical nature of
chirality has not been so far well recognized
and treated in mechanics. In
particular, the fact that it does not obey time invariance.
As shown before, this can be readily deduced by a simple
momentum consideration
for the operation of a rotating water
sprinkler. The vectorial velocity of the stream of water changes
its direction at the bending of the sprinkler, and the momentum
change caused by this is transferred to the sprinkler, which
determines the sense of rotation of the sprinkler (see Fig. 1).
Upon ``reversing''
time, the stream of water ``reverses'' its direction of flow.
Let us now analyze briefly the situation if the stream of water
reverses, its direction of flow. Let $\vec{v}_1$ and $\vec{v}_2$
be the velocity vectors of the original flow of water before and
behind the sprinkler bending, respectively, and
$\Delta \vec{v} = \vec{v}_1 - \vec{v}_2$
be the change in these velocities, which is proportional to the
momentum transfer to the sprinkler. Now let $\vec{u}_1$ and $\vec{u}_2$
be the opposite velocities, respectively, upon reversing the direction
of the water stream. It is obvious that
$\vec{u}_1 = - \vec{v}_2$
and
$\vec{u}_2 = - \vec{v}_1$.
Let
$\Delta \vec{u} = \vec{u}_1 - \vec{u}_2$,
then
\begin{eqnarray}
\Delta u = \vec{u}_1 - \vec{u}_2 = (-\vec{v}_2) - (- \vec{v}_1)
= \vec{v}_1 - \vec{v}_2 = \Delta \vec{v}  %1
\end{eqnarray}
which proves that the change of velocity at the sprinkler bending
remains the same regardless of the water flow reversal.  This is also
shown in detail in Fig. 1.
A similar situation exists also for the rotating radiometer.
No matter from which direction the light is shining, at its blade,
it does rotate always in the direction of the black
wing owing to the existing physical situation.
Such a behavior does
not obey time reversal
invariance, but rather space-time invariance.
Let us now express the space-time inversion by $S$ and $T$,
respectively: \ then a rotating water sprinkler does obey the
ST-invariance.
The same is true for all the examples given here of macro-chiral devices.
The same is also true for the protein
molecule example. A similar rule of PT-invariance (or CPT invariance) is
recognized in physics due to the presence of a spin in quantum mechanics,
but it is absent in classical physics because the concept
of physical chirality has been so far completely overlooked at.
In particular,
its mechanical aspects.
This concept of chirality appears much more in
chemistry due to the presence of many chiral molecules in organic
compounds, but chirality is mostly regarded and treated in chemistry
in terms of geometric shape, rather
than in its physical properties and contents. For this reason the concept
of chiral interaction has so far
been largely ignored in researches concerning chirality.

Such space-time symmetry operations for chiral devices may contain also a
certain aspect of practicality.  This is so in contrast to their
presence in the
domain of elementary particles in physics. From this view point any
time-reversible process is almost completely useless from any aspect of
practicality. For instance, any machine operation that produces
a certain function or object,
or any information transfer process, are completely time-irreversible.
These
include also biomolecular functions such as enzymatic activity and other
processes which are totally time-irreversible. For such reasons
of practicality,
the function of chiral devices, or molecules, may be of special significance
in comparison to  time reversible phenomena.

The next consideration involves the mode of selection, where only
{\bf one} direction of rotation is excited by chiral interaction,
whereas the opposite
direction remains largely inactive. Judging it from a thermodynamical
aspect, what is happening here is that only one half of the energy
that can be activated by the device is excited by CI,
whereas the other half
remains mostly inactive. On a molecular level this means that only
one half of the energy states of the system that are populated by
CI become active, whereas
the other half remain relatively empty. In other words,
the system does not readily
reach thermal equilibrium. This conclusion is of very substantial and
significant meaning for living systems because reaching
thermal equilibriuim practically means death.

Another way of looking at this effect is from the view-point of ergodicity.
This concept was introduced by Boltzmann about a century ago, and it
considers the mode of approaching thermal equilibrium of a single particle.
This is done as a
process of time average, instead of a statistical average
of a large ensemble of particles. In
view of this, the average velocity of such a particle, in any given
direction, approaches zero as a function of time. This is not the case
if, for instance, the average angular velocity of a
rotating water sprinkler is regarded
as a function of time. This is, actually, true for any effect of CI, when
averaged as a function of time. The selection of one direction of
motion out of two possible ones, which is typical of CI, makes its mode
of motion to become a non-ergodic entity, which again causes it to
avoid thermal equilibrium. This property of CI on a microscopic level
is, apparently, one of the most crucial advantages that chirality, or
CI, does contribute to molecular biology. It does postpone thermal
equilibrium, or death, for a considerable length of time, so that
the biological
function of these molecules can go on and not be affected by
approaching thermal equilibrium.

In this context it is significant to mention also Schr\"{o}dinger,
who became interested in the phenomenon of life and wrote a
book$^{15}$ ``What is Life?'' in 1944. His main conclusion in this
book was: ``It feeds on negative entropy''. This expression
becomes also quite appropriate for the function of physical
chirality and this is the reason for borrowing it as part of the
title of the present article. This is exactly what chiral
interaction is performing in its mode of selecting only one
direction of motion. Such a mode of activity reduces considerably
the entropy of the system.
In relation to the phenomenon of non-ergodicity it is also important
to mention its relevance to the process of evolution, which is
crucial in biology.
It is reasonable to deduce that systems that reach readily thermal
equilibrium never do undergo the process of evolution and remain
basically unchanged forever. Non-ergodic molecular systems have a
better chance to undergo evolutionary changes.
\end{section}
%
%\vspace{1cm}
%
%\noindent {\bf IV \  Discussion and Conclusions} \\
%
\begin{section}{Discussion and Conclusions}
Another aspect of chiral interaction regards the nature of this
effect, as well as the
size of energy that is involved in such a process. In discussing this
case it is not relevant to consider macro-chiral devices, and our main
concern is chiral interaction of biomolecular systems.
Unfortunately, our
knowledge at present, of this effect is quite limited and this is mainly
because of
the small amount of energy involved in this effect,
which is quite difficult to observe
experimentally. This may evoke criticism as to its possible significance.
Such a criticism is rather common among scientists who tend to attribute
significance to energy according to its size. What may be much more
significant than the amount of energy involved in living processes,
is its
quality, or degree of sophistication. This is particularly so in complex
systems such as certain biomolecules, proteins for example. The feature
of time-irreversibility of CI does contribute a degree of sophistication
to the activity of biomolecules.
In addition to this, there exist quite a few examples of
highly sophisticated
modes of energy which do require rather minute quantities of energy.
For instance,
an information transfer process requires a high degree of
sophistication in
wave modulation, but its size of energy is relatively small. In comparison,
boiling a kettle of water requires much more energy, but
what is its degree of sophistication? Another example is the small
amount of energy required to switch on and off a much larger source
of energy. This example can be regarded as a mode of
control mechanism energy,
which may also be the of significance to CI in biology.
Another, rather cruel example,
concerns the magnitude of energy change that occurs over a short time
interval during which a creature ceases to live. The change in energy
is rather small but its significance is very impressive. In these examples
and many others, the amount of energy involved in their performance
is of little interest, but their main effect is in their degree of
sophistication.

The experiment of Koumura et al.$^9$ is a significant and
important progress on a molecular level. This is the first
successful proof that a monodirectional molecular rotor can be
fabricated. The chiral interaction leading to this operation is
based on light radiation. This experiment supplies for a beginning
of support for the assumption of non-ergodic nature of chiral
interaction on a molecular level, being the basis for its negative
entropy and therefore for its avoiding readily thermodynamical
equilibrium. In addition to this there exists also the analysis of
behavior of globular proteins$^{5-7,10-12}$ which also react in a
monodirectional manner. \ The experiment of Careri et al.$^{13}$
provides for a supporting evidence for the significance of CI in
the enzymatic activity of proteins. It is quite reasonable to
assume that in biology, or in any living substance, the presence
of such modes of highly sophisticated and low energy signals may
have an important function in its life process.

In conclusion, let us mention again the significance and
importance of the phenomenon of chirality in biology. In
particular the features of chiral interaction (CI) that differ
largely from those of classical physics that do not contain chiral
structures in their interactions. These include the ST-invariance
of chiral interaction, which causes it to be time-irreversible.
The selectivity nature of CI by preferring one mode of motion out
of two possible ones, enables CI to become non-ergodic (or, more
specifically, to delay considerably its reaching thermal
equilibrium), which is a crucial element in life processes and
biological evolution. In other words: ``it feeds on negative
entropy'', as Schroedinger wrote in his book.
\end{section}
%
%\vspace{0.5cm}
%
%
%\noindent {\bf Acknowledgement}: \ \
%
\section*{Acknowledgement}
%\noindent
The author wishes to thank
Sharon and Yoram Yihyie for producing the figures in this article
as well as Gila Etzion for their help in completing this article.
%
%\pagebreak
%
%\begin{center}
%{\bf References}
%\end{center}
%
%\begin{enumerate}
%

%
%\end{enumerate}
%
%\vspace{1cm}
%
\end{document}